\documentclass[aps,reprint,10pt,showkeys,showpacs,superscriptaddress,showpacs]{revtex4-2}
\usepackage{multirow}
\usepackage{enumitem}
\usepackage{amsfonts} 
\usepackage{amsmath}
\usepackage{amssymb}
\usepackage{graphicx}
\usepackage{subfigure}
\usepackage{color}
\usepackage{enumerate}
\usepackage[]{natbib}
\usepackage{sidecap}
\usepackage{soul}
\usepackage{cancel}
\usepackage{ulem}
\usepackage{dcolumn}%
\usepackage{bm}%

\newcommand*\xbar[1]{%
  \hbox{%
    \vbox{%
      \hrule height 0.5pt 
      \kern0.5ex
      \hbox{%
        \kern-0.1em
        \ensuremath{#1}%
        \kern-0.1em
      }%
    }%
  }%
} 
\begin{document}

\title{Analytical and numerical study of plane-progressive thermoacoustic shock waves in complex plasmas}  
 \author{A. P. Misra}%
 \homepage{Author to whom any correspondence should be addressed}
 \email{apmisra@visva-bharati.ac.in; apmisra@gmail.com}
 \affiliation{Department of Mathematics, Siksha Bhavana, Visva-Bharati University, Santiniketan-731 235,  India
} 
\author{Gadadhar Banerjee}%
\email{gban.iitkgp@gmail.com. Current affiliation: Department of Mathematics, Burdwan Raj College, University of Burdwan, Burdwan 713 104, India. }
\affiliation{Department of Mathematics, Siksha Bhavana, Visva-Bharati University, Santiniketan-731 235,  India
} 

\date{\today}

\begin{abstract}
The formation of thermoacoustic shocks is  studied 
 in a  fluid complex plasma. The   thermoacoustic wave mode can be damped (or anti-damped) when the contribution from the thermoacoustic interaction is lower (or higher) than that due to the  particle collision and/or the  kinematic viscosity.  In the nonlinear regime,  the thermoacoustic wave, propagating with  the acoustic speed,   can evolve into small amplitude  shocks whose dynamics are governed by the Bateman-Burgers equation with  an additional nonlinear   term that appears due to the particle collision  and nonreciprocal interactions of charged particles providing the thermal feedback. The appearance of such nonlinearity  can cause the shock fronts to be stable (or unstable) depending on the collision frequency remains below (or above) a critical value and the thermal feedback is positive. The existence of different kinds of shocks and their characteristics  are   analyzed  analytically and numerically  with the system parameters that characterize the thermal feedback, thermal diffusion, heat capacity per fluid particle, the particle collision and the fluid viscosity. A good agreement between  analytical and numerical results is also noticed.
\end{abstract}


\maketitle
\section{Introduction} \label{sec-intro}
  Thermoacoustic instability \cite{nicoud2005,ghirardo2018} is important from  a fundamental point of view, as  it can cause pulsations leading to  their  accelerations in  specific environments, including  where the combustion reaction  occurs \cite{alekseev2001,kiverin2016,yurchenko2017}. Recently, the onset of such instability in complex plasmas has been reported in  the linear regime \cite{kryuchkov2018,garai2020}. In Ref. \cite{kryuchkov2018}, it was shown that the nonreciprocal effective interactions of charged particles can provide positive thermal feedback which, in turn,  leads to the amplification of thermoacoustic waves. The theoretical prediction  of such instability was also verified experimentally in that work. In another work \cite{garai2020}, the thermoacoustic instability has also been studied in the weakly and strongly coupled dusty plasma systems. The linear and nonlinear propagations of thermoacoustic waves and the nonlinear saturation of the thermoacoustic instability have been studied   in gas-filled tubes \cite{karpov2000,sugimoto2019}.  Furthermore, Navier-Stokes simulations have been  performed  for modelling of a theoretical travelling-wave thermoacoustic heat engine \cite{scalo2015}, as well as to identify the linear and nonlinear regimes of thermoacoustic  wave amplification and the generation of shocks in a minimal-unit looped resonator \cite{gupta2017}. On the other hand, the generation of shock waves has been explored in the thermoacoustic gas oscillations in a gas column with a high temperature gradient. It has been  observed that the periodic shocks can appear in the  travelling wave mode oscillations in which the temperature gradient acts as a source of the acoustic energy  \cite{biwa2011,biwa2014}.  However, the evolution of thermoacoustic shocks in complex plasmas has not been investigated  so far. 
\par 
The aim of this  letter  is to revisit the theory of 
 thermoacoustic waves, especially in the nonlinear regime, and to investigate the generation of thermoacosutic shocks propagating  at near-acoustic speeds as well as their characteristics in  fluid complex plasmas. Our analysis shows that the thermal feedback can    indeed induce shock waves   and they can be damped or anti-damped depending on whether the collision frequency is below or above a critical value. A good qualitative agreement of the analytical and numerical results, obtained, respectively,   by the tanh perturbation expansion scheme and   the Runge-Kutta scheme  is  also found to justify the existence of different kinds of  shocks.        
\section{Fluid model and dispersion relation} \label{sec-model-disp}
The equations describing the generation of   acoustic-like waves in one space dimension in a  fluid complex plasma with   the effects of temperature gradient, thermal feedback, the particle collision and the fluid kinematic viscosity are \cite{kryuchkov2018,alfaro1989} 
\begin{equation} \label{eq-cont}
\frac{\partial \rho}{\partial t}+\frac{\partial}{\partial x}\left( \rho v \right)=0,
\end{equation}
\begin{equation} \label{eq-moment}
\frac{\partial v}{\partial t}+v \frac{\partial v}{\partial x} = -\frac{1}{\rho }\frac{\partial }{\partial x}\left(\rho T\right) - \nu v + \mu \frac{\partial^2 v}{\partial x^2},
\end{equation}
\begin{equation} \label{eq-temp}
\frac{\partial T}{\partial t}+v \frac{\partial T}{\partial x} = \chi \frac{\partial^2 T}{\partial x^2}-\frac{2 \nu}{\Gamma}(T-1)+q,
\end{equation}
where  $\rho=nm$   is the fluid mass density normalized by its equilibrium value $\rho_0$ with $n$ denoting the number density and $m$ the mass,  $v$ is the  center-of-mass fluid flow velocity normalized by the acoustic speed $c_s =\sqrt{k_BT_0/m}$ with $k_B$ denoting the Boltzmann constant, $T$ is the total thermodynamic temperature normalized by its equilibrium value $T_0$, $\nu$ is the   collision frequency (damping rate) normalized by the plasma oscillation frequency $\omega_{p}=\sqrt{n_0 Q^2/\epsilon_0m}$ with $Q$ denoting the particle charge, $\mu$ is the coefficient of  fluid kinematic viscosity normalized by $\lambda^{2}_D \omega_{p}$ with $\lambda_D=c_s/\omega_{p}$ denoting the effective Debye length, $\chi$ is the coefficient of thermal diffusivity normalized by $c_s^2/\omega_{p}$, $\Gamma$ is the heat capacity, and $q(\rho,T)$ is the heat source normalized by $T_0\omega_{p}$. Also, the the space and time coordinates $x$  and $t$ are normalized, respectively,     by $\lambda_D$ and $\omega_{p}^{-1}$. It is to be mentioned that  Ref. \cite{kryuchkov2018} considered a model of two-dimensional complex fluids without the effects of kinematic viscosity \cite{alfaro1989}.  We are, however,   interested in the one-dimensional propagation of waves with the effects of   the   kinematic viscosity, which will contribute not only to the source of  damping of linear wave modes but also to the Burgers term in the nonlinear evolution equation of thermoacoustic   shocks. 
\par 
We  consider  the propagation of one-dimensional  planar compressional waves in complex plasmas along the $x$-direction. Assuming that the perturbations of density, velocity, and temperature are small compared to their equilibrium values and they vary as plane waves of the form $\exp(ikx-i\omega t)$ with the wave number $k$ and the wave frequency $\omega$, and linearizing Eqs. \eqref{eq-cont}-\eqref{eq-temp} about the equilibrium state with $T=1$, $\rho=1$ and $v=0$, we obtain the following  linear dispersion law \cite{kryuchkov2018}.
 \begin{multline} \label{eq-disp}
\left[\omega^2+ i \left( \nu+ \mu k^2\right)\omega   -k^2 \right]\\  
\times\left[ \omega+i \left( \chi k^2 + \frac{2 \nu}{\Gamma} -q_{T} \right) \right] 
=i q_{\rho} k^2,  
\end{multline}
where the parameters $q_{T}=\left( {\partial q}/{\partial T} \right)_0$ and $q_{\rho}=\left( {\partial q}/{\partial \rho} \right)_0$, calculated at   the equilibrium values  $T=1$ and $\rho=1$, correspond to the thermal feedback of the media to the temperature and density variations. Note that the wave frequency $\omega$ (or wave number $k$) becomes complex due to the effects of the thermal feedback as well as the collisional and the viscosity effects.  The dispersion equation \eqref{eq-disp} agrees exactly with that in Ref. \cite{kryuchkov2018} except the term proportional to
 $\mu$  due to   the effect of the kinematic viscosity.  In absence of the heat source or thermal feedback, i.e., for $q_{\rho}=q_{T}=0$,  the acoustic mode and the thermal mode (corresponding  to the first and second factors  in the square brackets of Eq. \eqref{eq-disp} respectively) are decoupled, in which case Eq. \eqref{eq-disp} gives
\begin{equation} \label{eq-disp1}
\omega=-i \left( \chi k^2 + \frac{2 \nu}{\Gamma}\right),
\end{equation}
and 
\begin{equation} \label{eq-disp2}
 \omega =k\sqrt{1-\frac{(\nu+k^2 \mu)^2}{4 k^2}}-i \frac{\nu+k^2 \mu}{2}. 
\end{equation}
Equation \eqref{eq-disp1} corresponds to a purely damped thermal mode whereas Eq. \eqref{eq-disp2} corresponds to an acoustic-like wave with the damping rate proportional to
 $(\nu+\mu k^2)/2$  due to the particle collision and/or the   kinematic viscosity, provided that $k$ lies in a very small interval $1-a<k<1+a$, where $a=\sqrt{1-\nu\mu}$.  This small regime is, however,  not of interest and thus inadmissible. So, we look for an acoustic mode with $\omega\simeq k+\delta \omega$, where  $\delta \omega$ is a small correction to the wave frequency due to the effects of particle collision,  the   kinematic viscosity,  and the thermoacoustic interactions. Separating the real and imaginary parts, from Eq. \eqref{eq-disp} we obtain the following expressions for the real wave frequency and the growth/damping rate \cite{kryuchkov2018} (for details, see Appendix \ref{appendix-B}). 
\begin{equation} \label{eq-omegaR}
\Re \omega \simeq k\left[ 1+\frac{q_{\rho}}{2} \frac{\chi k^2+2\nu / \Gamma-q_{T}}{k^2+\left(\chi k^2+2\nu / \Gamma-q_{T}\right)^2}\right],
\end{equation}
 \begin{equation} \label{eq-omegaI}
\Im \omega \simeq -\frac{1}{2}\left(\nu+\mu k^2\right)+ \frac{q_{\rho}}{2} \frac{k^2}{k^2+\left(\chi k^2+2\nu / \Gamma-q_{T}\right)^2}.
\end{equation}
From Eq. \eqref{eq-omegaR} it follows that the phase velocity of the thermoacoustic wave is constant, i.e., the wave becomes dispersionless in the long wavelength limit $k\rightarrow0$. However, as $k$ increases,   in contrast to the low-frequency acoustic waves,  the   phase velocity also approaches a constant value (i.e., close to the   acoustic speed $c_s$). On the other hand, from Eq. \eqref{eq-omegaI} it is evident that the wave can be stable (damped) or unstable (anti-damped) depending on whether the thermal contribution to the wave mode is smaller or larger than that associated with the collision and the viscosity. Typically, for positive values of $q_{\rho}$ and small values of $\nu$ and $\mu$, $\Im\omega$   becomes positive, implying the wave amplitude to grow, leading to the thermoacoustic instability \cite{kryuchkov2018}. 
\par
\begin{figure*}
\centering
\includegraphics[width=7in,height=3.5in]{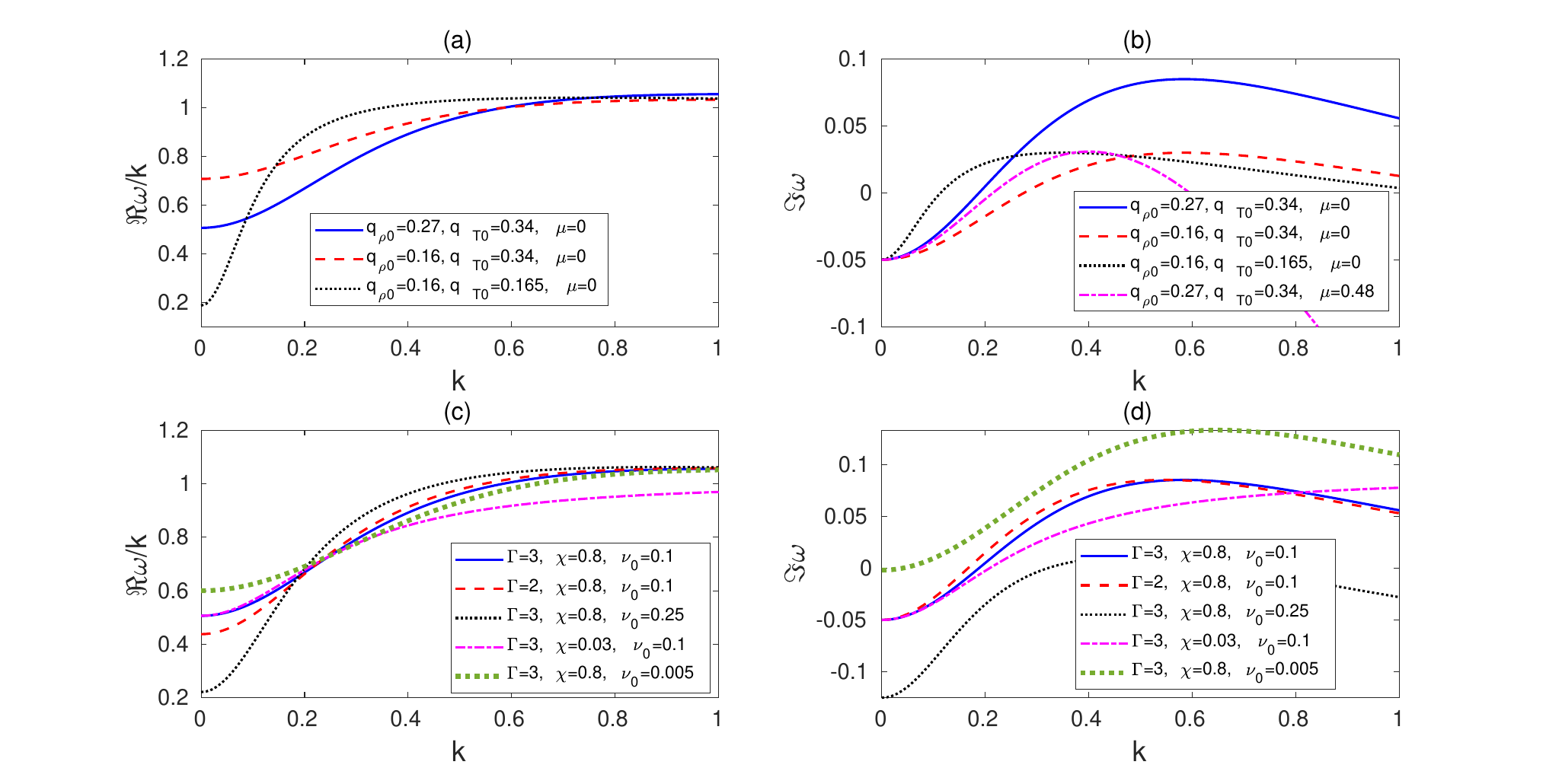}
\caption{ The  real (subplots (a) and (c)) and the imaginary (subplots (c) and (d))  parts of the wave frequency $\omega$ are plotted against the wave number $k$ for different values of the parameters as in the legends to show the instability growth and decay rates of the wave. The fixed parameter values for the subplots ((a), (b)) and ((c), (d)), respectively, are $(\Gamma=3,~\chi=0.8,~\nu_0=0.1)$ and $(q_{\rho_0}=0.27,~q_{T_0}=0.34,~\mu=0)$.}
\label{fig1-disp}
\end{figure*}

 
 In order to study the characteristics of the wave mode and the growth/decay rate in details, we numerically analyze Eqs. \eqref{eq-omegaR} and \eqref{eq-omegaI}. The results are displayed in Fig. \ref{fig1-disp}. The profiles of $\Re\omega$ are shown in subplots (a) and (c), and those of   $\Im\omega$ are in subplots (b) and (d) for different values of the parameters, namely $q_{\rho}$ and $q_{T}$ that characterize the thermal feedback; the heat capacity per particle $\Gamma$, the coefficient of the thermal diffusivity $\chi$, the collision frequency $\nu$, and the coefficient of the kinematic viscosity $\mu$. It is noted that although the wave frequency increases with $k$, the increase or decrease of the phase velocity depends on the ranges of values of $k$ as well as those of the parameters. Interestingly,  for $k\ll1$ and $k>1$, the phase velocity approaches a constant value. While in the former case, the phase velocity remains smaller than $c_s$, in the latter, it may slightly exceed the same. However, a domain of $k~(<1)$ exists corresponding to a wide range of values of the parameters where the phase velocity is close to the acoustic speed $c_s$.  Thus, in contrast to typical acoustic  waves (e.g., ion-acoustic wave), the thermoacoustic wave can propagate with sonic, subsonic or supersonic velocity. On the other hand, both the growth and the decay of the wave amplitude can occur in a wide range of values of $k$ and the parameters. In this context,  a purely growing mode or a purely damped mode can also exist when the collison frequency $\nu$ remains below or above a  critical value. For example, given $q_{\rho}=0.27$,  $q_{T}=0.34$, $\chi=0.8$, $\Gamma=3$ and $\mu=0$, $\Im\omega>0$ or $\Im\omega<0$, respectively when $\nu\lesssim0.005$ or $\nu\gtrsim0.25$ (see the thick and thin dotted lines of subplot (d)). Physically, in absence of the fluid viscosity, when the particle's collision frequency is below (above) a critical value, its contribution to the wave frequency ($\Im\omega$) becomes smaller (larger) than that due to the thermal feedback of the medium. As a result, only the instability (damping) is seen to occur.    It is also important to note that when the viscosity is absent $(\mu=0)$ in the medium, the growth/decay rate in both the cases  reaches a maximum value within the domain $0<k<1$ having a cutoff at a higher $k>1$ except the case with nonzero $\mu$ (see the dash-dotted line in subplot (d)).     
\section{Nonlinear evolution of shocks}\label{sec-nonl}
 In Sec. \ref{sec-model-disp},  we have seen that depending on the parameter values, the thermoacoustic interaction of density and temperature perturbations can  lead to either the wave instability or     damping. In this section,  we consider this interaction effect   and examine whether   the    linear perturbations, as they propagate and  the nonlinear effects intervene, can develop into nonlinear compressive or rarefactive shocks with oscillatory or monotonic profiles. To this end, we derive an evolution equation for  small amplitude shocks and study their characteristics in the physical parameter space.     
 \par 
 In what follows, we are interested in the evolution of one-dimensional plane progressive shocks in a frame that moves along the $x$-axis with a velocity close to the acoustic speed $c_s$. The wave can achieve this speed at a finite value of $k$ such that $k$ is not too small but can be smaller than  unity and the parameters satisfy the condition $q_\rho/\left(4\nu/\Gamma-2q_T\right)<1$ (\textit{cf}. subplots (a) and (c) of Fig. \ref{fig1-disp}).     Thus, in our reductive perturbation scheme  the new space and time coordinates can be defined as  \cite{cousens2014} (for details, see Appendix \ref{appendix-B})
\begin{equation} \label{eq-stretch}
\xi =\varepsilon   \left(x-  t\right) ~\text{and }\tau = \varepsilon ^2 t,
\end{equation}
where $\varepsilon$ is an expansion parameter that reflects the smallness of wave perturbations. We further assume that $\nu\sim\nu_0 \varepsilon$, $q_{\rho}\sim  q_{\rho_0}\varepsilon$, and $q_{T}\sim  q_{T_0}\varepsilon$, where $\nu_0,~q_{\rho_0}$, and $q_{T_0}$ are each of the order of unity.
The dependent variables are expanded as 
\begin{multline} \label{eq-expansion}
\rho = 1+\varepsilon \rho_1+\varepsilon^2 \rho_2+\varepsilon^3 \rho_3+ \cdots,  \\ 
 v = \varepsilon v_1+\varepsilon^2 v_2+\varepsilon^3 v_3+ \cdots,    \\
 T = 1+\varepsilon T_1+\varepsilon^2 T_2+\varepsilon^3 T_3+ \cdots,   \\
q(\rho,T) = \varepsilon^2  q_1+\varepsilon^3 q_2+ \cdots,  \\
\end{multline}
where $q_1\equiv q_{\rho_0} \rho_1+q_{T_0} T_1$, $q_2\equiv q_{\rho_0} \rho_2+q_{T_0} T_2 $ etc.     are obtained by evaluating the first and higher order partial derivatives  of $q(\rho,T)$ with respect to $\rho$ and $T$ at their equilibrium values From Eq. \eqref{eq-expansion}, it is to be noted that while the first order perturbation in each of $\rho$, $v$ and $T$ scales as $\epsilon$, that of $q$ scales as $\epsilon^2$ due to its Taylor series expansion about the equilibrium values of $\rho$ and $T$  (for details,   see Appendix \ref{appendix-A}).  
\par  
Next, we substitute the new stretched coordinates from Eq. \eqref{eq-stretch} and the expansion from Eq. \eqref{eq-expansion} into Eqs. \eqref{eq-cont}-\eqref{eq-temp}, and equate the coefficients of different powers of $\varepsilon$.  
In the lowest order  of $\varepsilon$,   we obtain the following expressions for the first order perturbations.
\begin{equation} \label{eq-pert1}
v_1=\rho_1, ~~T_1=-\frac{\Gamma^{\prime}}{2} \rho_1,
\end{equation}
where $\Gamma^{\prime}=2\left( \nu_0-q_{\rho_0} \right)/\left(2 \nu_0 / \Gamma-q_{T_0}\right)$. 
\par 
 From the next order of $\varepsilon$, we obtain the following expressions for the second order perturbations in terms of first order quantities.
\begin{equation} \label{eq-pert2a}
\frac{\partial v_2}{\partial \xi}-\frac{\partial \rho_2}{\partial \xi}=-\frac{\partial \rho_1}{\partial \tau}-2 \rho_1 \frac{\partial \rho_1}{\partial \xi},
\end{equation}
\begin{equation} \label{eq-pert2b}
\frac{\partial T_2}{\partial \xi}-\frac{\partial v_2}{\partial \xi}+\frac{\partial \rho_2}{\partial \xi}=-\frac{\partial \rho_1}{\partial \tau}+\Gamma^{\prime} \rho_1 \frac{\partial \rho_1}{\partial \xi}+ \mu \frac{\partial^2 \rho_1}{\partial \xi^2}-\nu_0 \rho_1^2,
\end{equation}  
\begin{equation} \label{eq-pert2c}
-\frac{\partial T_2}{\partial \xi}=-\frac{\Gamma^{\prime}}{2} \chi \frac{\partial^2 \rho_1}{\partial \xi^2}+\frac{\Gamma^{\prime}}{2} \frac{\partial \rho_1}{\partial \tau}+\frac{\Gamma^{\prime}}{2} \rho_1 \frac{\partial \rho_1}{\partial \xi}.
\end{equation}
\par
After eliminating the second order  perturbed quantities from  Eqs. \eqref{eq-pert2a}-\eqref{eq-pert2c}   and using the   results, given by,  Eq. \eqref{eq-pert1}, we obtain   the following   Bateman-Burgers  or simply the Burgers-like equation  for thermoacoustic shocks in fluid complex plasmas.
\begin{equation} \label{eq-Burgers}
\frac{\partial \rho}{\partial \tau}+A \rho \frac{\partial \rho}{\partial \xi}=B \frac{\partial^2 \rho}{\partial \xi^2}+D \rho^2,
\end{equation}
where $\rho(\xi,\tau)\equiv \rho_1(\xi,\tau)$ and the coefficients of the nonlinear convection, the diffusion and the   nonlinearity associated with the particle collision and the thermal feedback (Hereafter, we call it as ``collisional nonlinearity''), respectively, are 
\begin{equation} \label{eq-coeff}
A=\frac{3 \Gamma^{\prime}-4}{\Gamma^{\prime}-4},~~B=\frac{\Gamma^{\prime} \chi-2 \mu}{\Gamma^{\prime}-4},~~D=\frac{2 \nu_0}{\Gamma^{\prime}-4}.
\end{equation}
In deriving Eq. \eqref{eq-Burgers}, we have neglected the secular terms involving $-\nu_0v_2$ (on the right-hand side of   Eq. \eqref{eq-pert2b}) and $q_2-2\nu_0T_2/\Gamma$ (on the right-hand side of   Eq. \eqref{eq-pert2c}) by assuming $\partial/\partial\xi\gg1/L$, where $1/L=\max\left\lbrace2\nu_0/\Gamma-q_{T_0}, ~q_{\rho_0},~\nu_0 \right\rbrace$, since we look for a solitary shock solution of Eq. \eqref{eq-Burgers} for the first order density perturbation.     
\par 
We note that the nonlinear term proportional to 
 $D$ in Eq. \eqref{eq-Burgers} appears due not only to the particle collision but also to a finite value of the heat capacity per particle and is modified by the thermal feedback of the medium to the density and thermal fluctuations. Furthermore, the nonlinearity in $\rho$ appears due to the first order smallness of $\nu\sim \nu_0\varepsilon$. However, a second order smallness of $\nu\sim\nu_0\varepsilon^2$ could result in a linear term proportional to
 $\rho$ in Eq. \eqref{eq-Burgers}. So, the evolution dynamics of shocks with the linear term $D\rho$ should differ from  those with the nonlinear term $D\rho^2$.  Before going into the  dynamics of thermoacoustic shocks, it is imperative to investigate the nature of the coefficients $A$, $B$ and $D$  of Eq. \eqref{eq-Burgers} for different values of the parameters $q_{\rho_0}$, $q_{T_0}$, $\nu_0$, $\Gamma$, $\chi$ and $\mu$. The latter two, however, modify    the coefficient of diffusion $B$ only.  Such an investigation  is crucial not only to examine the existence of different kinds of shock solutions of Eq. \eqref{eq-Burgers}  that the   plasma medium can support but also to study their characteristics with the variation of parameters. Typically, the specific heat ratio $\Gamma$ ranges in between $2$ to $4$. Also,  we have $q_{\rho_0} \sim 0-0.4$, $q_{T_0} \sim 0-0.4$, $\chi \sim 0-1$, $\mu \sim 0-1$ and $\nu_0 \sim 0-1$ such that the condition $q_\rho/\left(4\nu/\Gamma-2q_T\right)<1$,  as stated before,  is fulfilled.
\par
\begin{figure*}
\centering
\includegraphics[width=6in,height=3.0in]{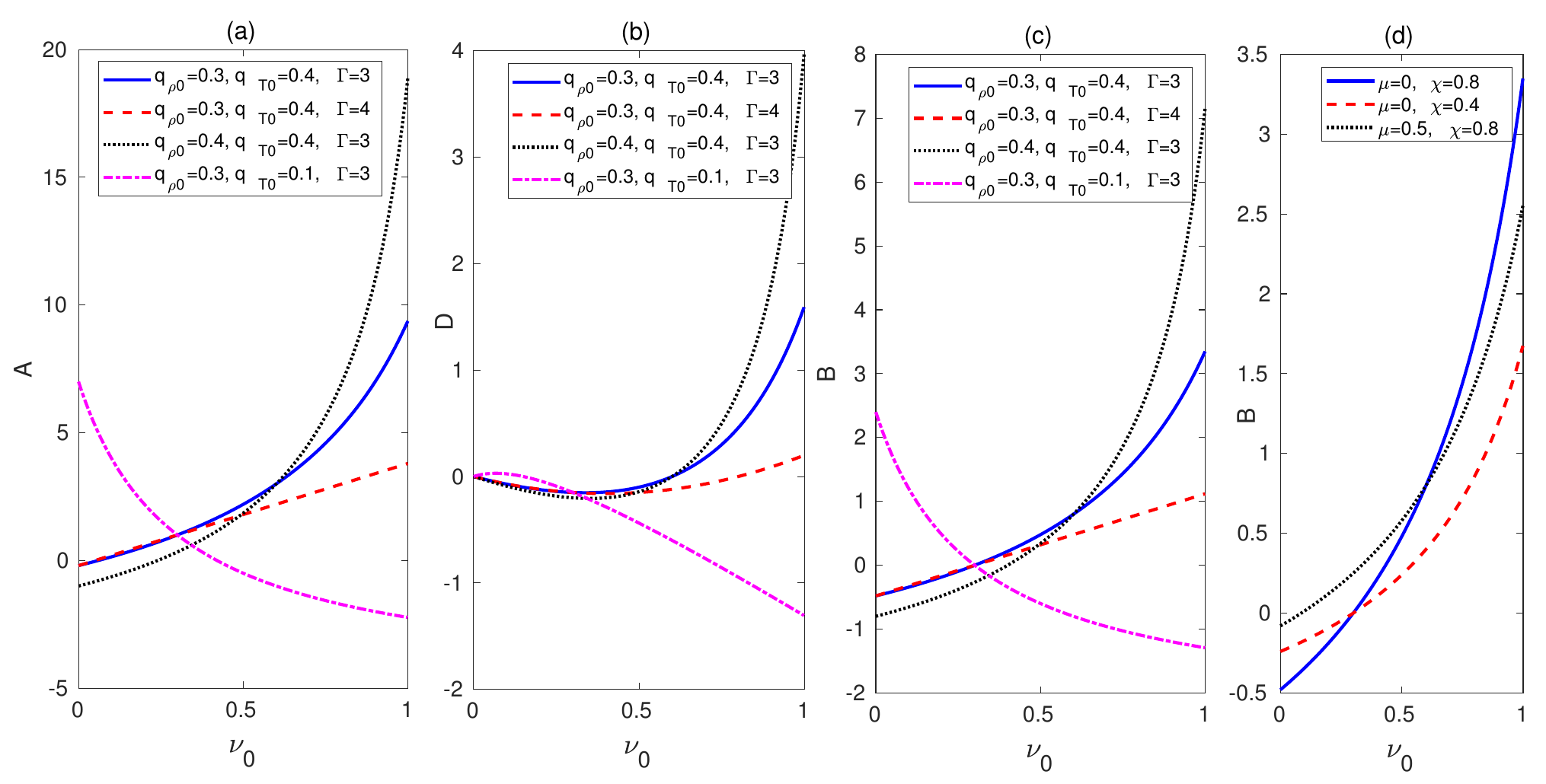}
\caption{The variations of the coefficients of Eq. \eqref{eq-Burgers} are shown in a domain of the collision frequency $\nu_0$ for different values of the parameters that characterize the heat capacity $(\Gamma)$, the thermal feedback due to density $(q_{\rho_0})$ and temperature $(q_{T_0})$ fluctuations, the thermal diffusion $(\chi)$ and the kinematic viscosity $(\mu)$ as in the legends.  The  fixed parameter values for the subplots [(a), (b), (c)] and (d), respectively, are $(\chi=0.8,~\mu=0)$ and $(q_{\rho_0}=0.3,~q_{T_0}=0.4,~\Gamma=3)$.}
\label{fig2-ABD}
\end{figure*}

 
 Figure \ref{fig2-ABD} displays the profiles of $A$, $B$ and $D$ with respect to the collision frequency $\nu$ for different values of the other parameters. Here, we note that while the sign  of $A$ could be crucial for determining the existence of compressive or rarefactive schocks, that of $D$ may characterize  the emergence of instability or damping of nonlinear shocks. Furthermore,  the magnitudes of $A$, $B$ and $D$ should be finite in order to avoid any blow  up solution of shocks.  It is found that all the coefficients can assume exceedingly high values when the collision frequency outstrips its critical value  $(\nu_c\sim0.5)$ and the thermal feedback to the temperature variation $(q_{T_0})$ is relatively high (see the solid, dashed and dotted lines in subplots (a), (b) and (c)). In this case,  the wave steepening can occur with  an increased value of the wave amplitude due to a dominating role of the nonlinear convection $(A)$ over the thermal diffusion and/or the kinematic viscosity $(B)$ such that the thermoacoustic shock wave may be unstable due to positive values of the nonlinear coefficient $D$ associated with the particle collision and the thermal feedback. However, when the value of $q_{T_0}$ is relatively low (see the dash-dotted lines in subplots (a), (b) and (c)), the coefficients $A$, $B$ and $D$ can be positive or negative depending on the ranges of values of $\nu_0$. Here, the magnitudes of   $A$, $B$ and $D$ decrease with increasing values of $\nu_0$ such that  $|A/B|>1$  in $0<\nu_0\lesssim0.3$ and $|A/B|\lesssim1$  in $0.3<\nu_0\lesssim1$. Furthermore, $D>0~(<0)$ in $0<\nu_0\lesssim0.1$ $(0.1<\nu_0\lesssim1)$. Thus, it follows that when the effects of the thermal feedback of the medium to the temperature variation ($q_{T_0}$) and the particle's collision   $(\nu_0)$ are relatively small, the   nonlinear convection still dominates  over   the thermal diffusion  (and/or the kinematic viscosity). As a result,  the wave instability may occur for the collsion frequency well below the plasma oscillation frequency, i.e., $\nu_0\lesssim0.1$.   However,   for frequencies in the interval $0.1<\nu_0\lesssim1$, the effects of the thermal diffusion and/or the kinematic viscosity become  comparable to that of the convective nonlinearity, and because of $D<0$, the shock waves can be damped.  On the other hand, $B$ tends to assume positive values in the entire domain of $\nu_0$ with an increasing value of the viscosity parameter $\mu$. From the subplots (a) and (c), it is observed that in the absence of the fluid viscosity and for a fixed value of $\chi$, which characterizes the thermal diffusivity, the qualitative features of $A$ and $B$ are almost the same except for their changes of signs and/or magnitudes with the variations of the parameters $q_{\rho_0}$, $q_{T_0}$ and $\Gamma$. From the profiles of $A$, $B$ and $D$ it may be concluded that in order to have a finite shock solution the values of $\nu_0$ may be restricted to lie in the interval   $0<\nu_0\lesssim0.5$. In the following section \ref{sec-sol}, we will consider these parameter regimes to investigate different kinds of shock solutions both analytically and numerically.  
\section{Shock solution: analytical  and numerical approach} \label{sec-sol}
We employ the tanh expansion scheme \cite{malfliet1993}  to obtain an approximate shock solution of Eq. \eqref{eq-Burgers}  when the coefficient $D$ of collisional nonlinearity is relatively small compared to the nonlinear convection and the diffusion, i.e., $|D|\ll |A|,~|B|$. We also numerically investigate the existence of different kinds of  shock solutions  and examine whether the qualitative features in both the cases agree in the same parameter space.  
\subsection{Analytical approach} \label{sec-analytic}
 We note  that in absence of the particle collision $(\nu=0)$, Eq. \eqref{eq-Burgers} reduces to the   known Burgers equation   whose asymptotic shock solution is   given by 
\begin{equation} \label{eq-sol-shock0}
 \rho(\xi, \tau)=\frac{v_0}{A} \left[1-\tanh\left\lbrace \left(\xi-v_0 \tau\right) \frac{v_0}{2B} \right\rbrace \right],
 \end{equation} 
where $v_0$ is the speed of the shock front and the  imposed  boundary conditions are $\rho, ~ \rho_\xi, ~ \rho_{\xi\xi}\rightarrow0$ as $\xi \rightarrow \infty$. Such a shock profile has two layers which may be  composed of  compressive and/or    rarefactive  wave fronts.  From Eq. \eqref{eq-sol-shock0}, it is clear that the shocks can be compressive $(\rho>0)$ or rarefactive $(\rho<0)$ according to when the coefficient of the nonlinear convection $A$ is positive or negative. We, however, skip the further analysis of this particular solution rather we look for a solution of Eq. \eqref{eq-Burgers} when the collisional nonlinearity  is no longer negligible but small compared to those associated with $A$ and $B$, i.e.,    $|D|\ll|A|,~|B|$. The case  with $|D|\gtrsim|A|,~|B|$ may result into blow up solutions which are physically inadmissible. 
\par 
It is to be noted while various conservation laws  (e.g., the conservation of total number of particles and  the conservation of energy) hold for the Burgers equation with $\nu=0$, the same do not hold for Eq. \eqref{eq-Burgers}    because of the nonlinearity  proportional to $D$. So, either the wave energy grows leading to the instability or decays to exhibit the damping. It follows that a steady state solution with finite wave energy of Eq. \eqref{eq-Burgers} does not exist  and thus,  we look for  an approximate solution  of it. To this end we recast  Eq. \eqref{eq-Burgers} as 
\begin{equation} \label{eq-Burgers2}
\frac{\partial \rho}{\partial \tau}+A \rho \frac{\partial \rho}{\partial \xi}-B \frac{\partial^2 \rho}{\partial \xi^2}+\lambda \rho^2=0,
\end{equation}
where $\lambda=-D$.   We employ the tanh perturbation expansion scheme   in which Eq. \eqref{eq-Burgers2} can be treated as a perturbed equation  with the small perturbation being proportional to $\lambda$. So, it is reasonable to assume Eq. \eqref{eq-sol-shock0} as the unperturbed solution of Eq. \eqref{eq-Burgers2} and   a slow time dependence of the wave amplitude and velocity of the new solution due to the nonlinearity  proportional to $\lambda$. Thus, for a constant $V$, we define a new transformation $\zeta$, retaining $\tau$ as is, as  
\begin{eqnarray} \label{eq-trans1}
\zeta=\frac{V}{B} \left[\xi-\phi(\tau)\right],   
\end{eqnarray}
so that  $\rho(\xi, \tau) \equiv \rho(\zeta, \tau)$  represents a localized solution of Eq. \eqref{eq-Burgers2} that travels with a velocity $d \phi /d \tau$ having the characteristic width $W=B/V$, which plays the role of the wavelength. Here, we do not consider any time dependency of $V$ in order to avoid any secular term proportional to
 $\xi$.   Using the transformation  \eqref{eq-trans1},  Eq. \eqref{eq-Burgers2} reduces to
\begin{equation} \label{eq-Burgers3}
\frac{\partial \rho}{\partial \tau}-\frac{V}{B} \frac{d \phi}{d \tau} \frac{\partial \rho}{\partial \zeta}+\frac{V A}{B} \rho \frac{\partial \rho}{\partial \zeta}-\frac{V^2}{B} \frac{\partial^2 \rho}{\partial \zeta^2}+\lambda \rho^2=0.
\end{equation}
\par 
Next, we look for a solution  in analogy with that of the  Burgers equation with $\lambda=0$ and use the similar boundary conditions, namely $\rho, ~ \partial\rho / \partial \zeta, ~ \partial^2 \rho / \partial \zeta^2 ~ \rightarrow0$ as $\zeta \rightarrow \pm\infty$. To this end, 
we introduce a variable $Y=\tanh(\zeta)$ for an infinite series expansion with $\tau$-dependent coefficients and assume  $\rho(\zeta, \tau)=S(Y,\tau)$. Thus,  Eq. \eqref{eq-Burgers3} gives
\begin{multline} \label{eq-Burgers4}
\frac{\partial S}{\partial \tau}+\lambda S^2 + \frac VB \left(1-Y^2\right)\left[ \frac{\partial S}{\partial Y}\left(-\frac{d \phi}{d \tau}+A S+2V Y\right)\right.  \\ 
\left. -V \left(1-Y^2\right)\frac{\partial^2 S}{\partial Y^2}\right]=0.  
\end{multline}
Based on the exact solution (Eq. \eqref{eq-sol-shock0}) of the conserved Burgers equation, we presume,  as an ansatz, that  the solution of Eq. \eqref{eq-Burgers4} will take the form
\begin{multline} \label{eq-ansatz1}
S(Y, \tau)=G(\tau) (1-Y) \left[ 1+a_1(\tau) Y + a_2(\tau) Y^2 \right.   \\
\left.+ a_3(\tau) Y^3 + a_4(\tau) Y^4 + a_5(\tau) Y^5 + \cdots \right],  
\end{multline}
where $G(\tau)$ and $a_i(\tau)$, for $i=1,2,3,...$, are unknown functions of $\tau$, and the series in $Y$ is to be convergent.
\par 
 Substituting the ansatz \eqref{eq-ansatz1}   into Eq. \eqref{eq-Burgers4} and equating the coefficients of $Y^n$, for $n=0,1,2,...$,   to zero one can obtain different expressions for $G(\tau)$, $\phi(\tau)$ and $a_i(\tau)$ with $i=1,2,3,...$. First, to obtain the expressions for $G(\tau)$ and $\phi(\tau)$, we set $a_1(\tau)=a_2(\tau)=0$. The lowest order of $Y$ then  gives
\begin{equation} \label{abc1}
-\frac{V A}{B} G^2(\tau)+\lambda G^2(\tau)+G^{\prime}(\tau)+\frac{V}{B} \frac{d \phi}{d \tau} G(\tau)=0.
\end{equation}
Next, looking for a solution of Eq. \eqref{abc1} for $G(\tau)$ that decays with time $\tau$ and hence  the velocity $d\phi/d\tau$, we have
\begin{equation} \label{eq-G-tau1}
\frac{dG(\tau)}{d \tau}=-\lambda G^2(\tau),
\end{equation}
\begin{equation} \label{eq-phi-tau1}
\frac{d \phi}{d \tau}=A G(\tau).
\end{equation}
Using the initial conditions $G(\tau)=2V/A$ and $\phi(\tau)=0$ at $\tau=0$, and in analogy with the solution  of the conserved Burgers equation, we obtain for $\lambda\neq 0$ the following solutions of   Eqs. \eqref{eq-G-tau1}   and  \eqref{eq-phi-tau1}.   
\begin{equation} \label{eq-G-tau2}
G(\tau)=\left(\lambda \tau+\frac{A}{2V}\right)^{-1},
\end{equation}
\begin{equation} \label{eq-phi-tau2}
\phi(\tau)=\frac{A}{\lambda} \log\left(1+\frac{2 V \lambda}{A} \tau\right),
\end{equation}
In particular, for $\lambda=0$, Eqs. \eqref{eq-G-tau1}   and  \eqref{eq-phi-tau1} give
\begin{equation} \label{eq-G-tau3}
G(\tau)=\frac{2V}{A},
\end{equation}
\begin{equation} \label{eq-phi-tau3}
\phi(\tau)=v_0 \tau,
\end{equation}
where $v_0=2V$. Here, for $\lambda=0$, the exact solution can be written in terms of $S(Y,\tau)\equiv S(Y)$, i.e.,  
\begin{equation} \label{eq:solution_0}
S(Y)=\frac{v_0}{A}(1-Y),
\end{equation}
where $Y=\tanh(\zeta)=\tanh \left[(v_0/2B)(\xi-v_0 \tau)\right]$ and  the required boundary condition  is $S(Y)\rightarrow 0$ as $Y\rightarrow 1$. The solution \eqref{eq:solution_0} completely agrees with the solution \eqref{eq-sol-shock0} of the conserved Burgers equation.  From Eqs. \eqref{eq-phi-tau1} and \eqref{eq-G-tau2}  we note  that     both the  amplitude $G(\tau)$  and  the velocity   $d\phi/d\tau$ of the thermoacoustic shock may either    decay (damping) or grow (instability) with time depending on the values of the collision frequency below or above a critical value and those of $q_{\rho_0}$ and $q_{T_0}$. Furthermore, depending on these values of the parameters,  $G(\tau)$  can be either positive or negative implying the existence of either compressive or rarefactive shocks.  
\par 
 Next, the $\tau$-dependent quantities $a_i(\tau)$  in the ansatz \eqref{eq-ansatz1}  are obtained successively from the higher orders of $Y$ as follows:  
\begin{equation} \label{eq:a3}
a_3(\tau)=-\frac 13+\frac{1 }{6 V^2} (A V-B \lambda) G(\tau),
\end{equation}
\begin{equation} \label{eq:a4}
a_4(\tau)=a_3(\tau)+\frac{B \lambda}{12 V^2} G(\tau),
\end{equation} 

\begin{multline} \label{eq:a5}
a_5(\tau)=- \frac{8}{15} +\frac{1}{12 V^2} (4A V-3B \lambda)G(\tau)   \\
+ \frac{A}{30 V^3} \left(-AV+B \lambda\right) G^2(\tau),  
\end{multline}
\begin{multline} \label{eq:a4}
a_6(\tau)=a_5(\tau)+\frac{B \lambda}{9 V^2} G(\tau)   \\
 +\frac{B \lambda}{360 V^4}(-9AV+4B \lambda)G^2(\tau), 
\end{multline}
and so on. Finally, we obtain the following approximate shock solution of Eq. \eqref{eq-Burgers}.
\begin{multline} \label{eq-sol-shock1}
\rho(\xi,\tau)=\left(\lambda \tau +\frac{A}{2V}\right)^{-1} \left(1-Y\right)\\
\times \left[ 1+ a_3(\tau) Y^3+a_4(\tau) Y^4+\cdots\right], 
\end{multline}  
where
\begin{equation}
 Y=\tanh\left[ \frac{V}{B}\left( \xi-\phi(\tau)\right) \right].  
\end{equation}
It can be shown that for a particular choice of the  parameters, the condition $\vert a_{n+1}/a_n \vert < 1$ is satisfied so that the series in $Y$ converges. Here,  $a_n$ denotes the $n$-th term of the series in $Y$. In Eq. \eqref{eq-sol-shock1}, how many correction terms involving $a_i(\tau)$ are to be considered depends on the degree of smallness of the  perturbation proportional to
 $\lambda$. Typically, $|Y|\lesssim1$ and  for the parameter values as in Fig. \ref{fig4-analytic}(a) (solid line), we have $D\sim-0.012$; $a_3=-0.58$, $a_4=-0.34$, $a_5=-0.27$, and $a_6=-0.013$ at $\tau=200$ such that $|a_6(\tau)Y^6|\lesssim0.01$.  So, to the second order smallness of perturbation $(\sim0.01)$, we retain the correction terms upto $a_6(\tau)$. However, one can consider additional correction terms (beyond $a_6(\tau)$)  for more accuracy. In particular,  for $\lambda=0$, we have $a_i(\tau)=0$ with $i=3,4,5,..$, for which the solution   \eqref{eq-sol-shock0} is recovered.  As said before, if the smallness of the collision frequency is taken to be a higher order of $\varepsilon$     than the first order, i.e.,  $\nu\sim{\cal O}(\varepsilon^2)$, then the  term proportional to $D$ in Eq. \eqref{eq-Burgers}   appears
 as $D\rho$ instead of $D\rho^2$. In that case,    the shock solution can have either exponential growth or exponential decay   \citep{malfliet1993} which is significantly different from the present solution \eqref{eq-sol-shock1}.  
\par
\begin{figure*}
\centering
\includegraphics[width=7in,height=2.8in]{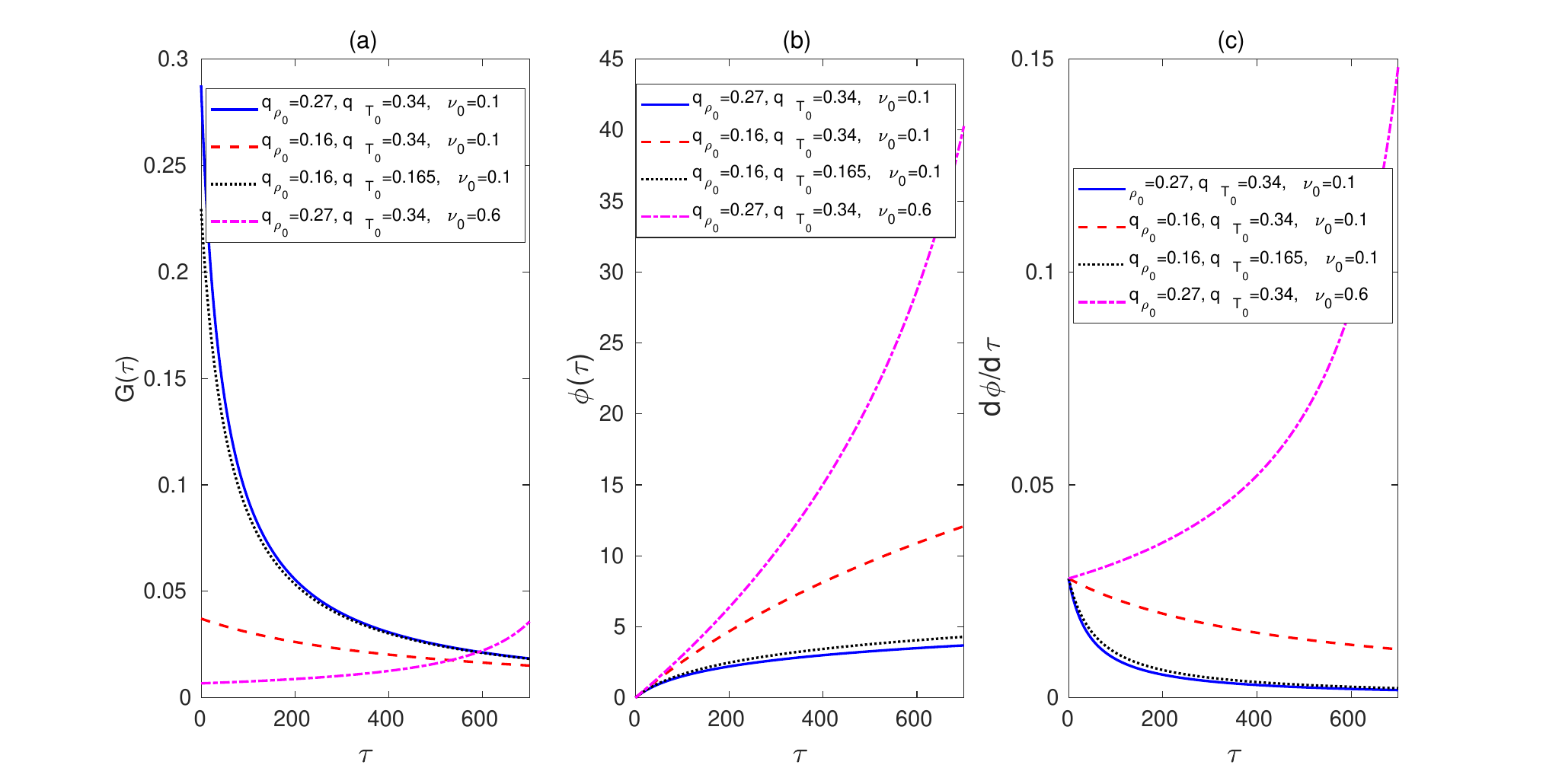}
\caption{The profiles  of the amplitude $(G(\tau))$, the relative propagation distance $(\phi(\tau))$ and the velocity $(d\phi/d\tau)$ of thermoacoustic shocks are shown for   fixed values of  $\Gamma=3$ and $V=1$, and  for different values of the other parameters as in the legends.  While the  solid, dashed and dotted lines correspond to the wave damping, the dash-dotted lines that to the instability of thermoacoustic shocks. }
\label{fig3-avd}
\end{figure*}

\begin{figure*}
\centering
\includegraphics[width=6in,height=3.0in]{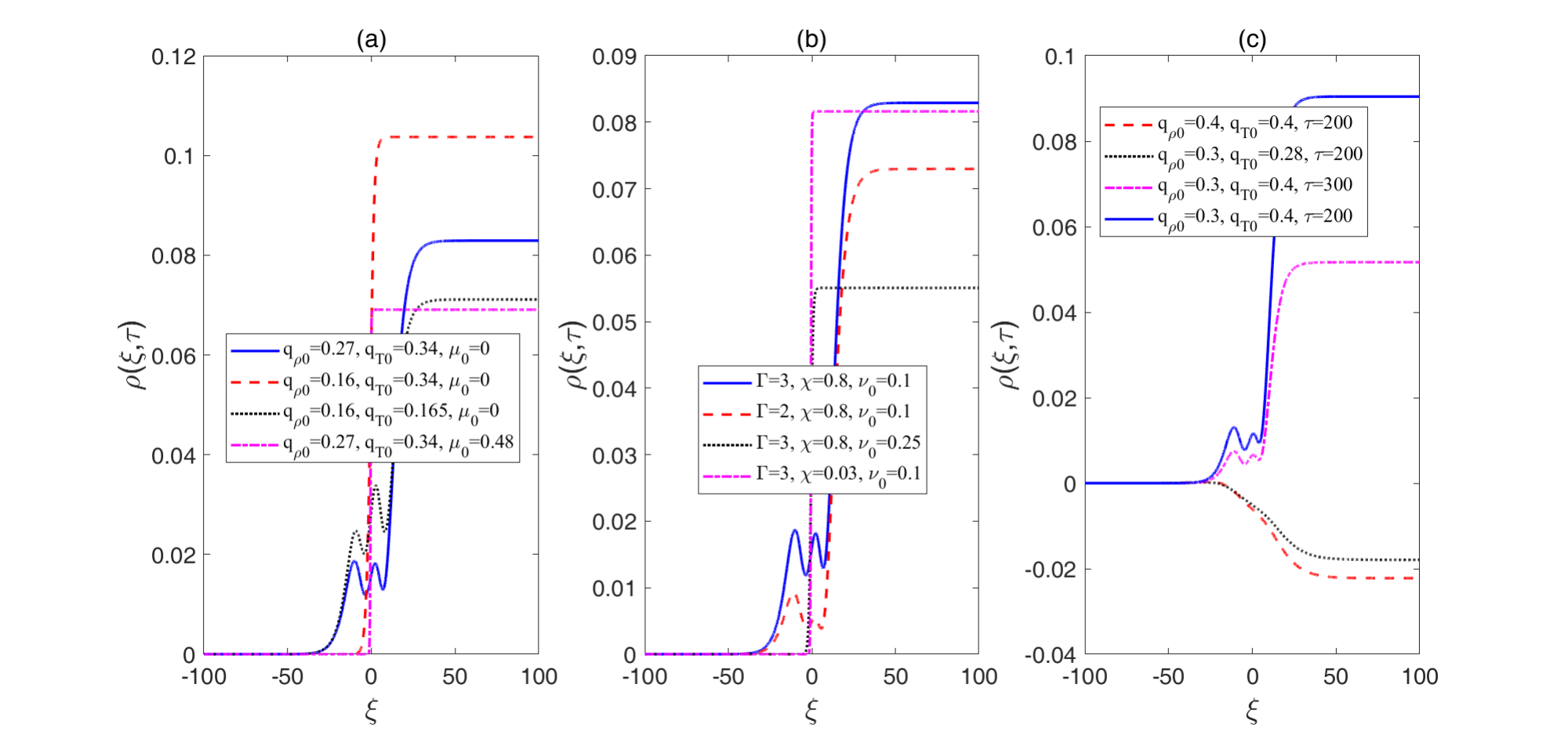}
\caption{The profiles  of the asymptotic shock solution \eqref{eq-sol-shock1} are shown with the variations of parameters as in the legends. The transition from oscillatory to monotonic shocks occur due to the effects of (a) the kinematic viscosity $(\mu)$ as well as the thermal feedback associated with density $(q_{\rho_0})$ and thermal $(q_{T_0})$ fluctuations and (b) the thermal diffusion $(\chi)$ and the particle collision $(\nu_0)$. The fixed parameter values for the subplots (a) and (b), respectively, are $(\Gamma=3,~\chi=0.8,~\nu_0=0.1)$ and $(q_{\rho_0}=0.27,~q_{T_0}=0.34,~\mu=0)$. Also, fixed are $\tau=200$ and $V=0.014$. Subplot (c) shows the existence of rarefactive shocks for a different set of parameter values as in the legend. The fixed parameter values in this case are $\Gamma=3$, $\nu_0=0.1$ and $\mu=0$.  }
\label{fig4-analytic}
\end{figure*}

 
Before we analyze the characteristics of the shock solution, given by,  Eq. \eqref{eq-sol-shock1}, it is important to investigate the properties of the amplitude $G(\tau)$ and the velocity $d\phi/d\tau$ of the shock as well as the relative propagation distance $\phi(\tau)$ for a fixed time at which the shock may eventually damp away. The results are displayed in Fig. \ref{fig3-avd}. It is evident that given fixed values of $\Gamma$, $q_{\rho_0}$ and $q_{T_0}$ when the collision frequency $\nu_0$ is below a critical value $\nu_c$, both the nonzero amplitude  and velocity of shocks decay with time (solid, dashed and dotted lines of subplots (a) and (c)). Since the  coefficient $D$ of the collisional nonlinearity (proportional to $\nu_0$) assumes small negative values for $\nu_0\lesssim\nu_c$ (\textit{cf}. Fig. \ref{fig2-ABD}), such a decay of the wave amplitude may result in the   damping of thermoacoustic shocks. Furthermore, in contrast to the  decay rate of the velocity (subplot (c)), the   decay rate of the wave amplitude can be lower (higher) with decreasing values of $q_{\rho_0}$ $(q_{T_0})$. Thus, the decay of the wave amplitude can be slowed down if the thermal feedback of the medium to the density fluctuation is typically low but that to the temperature fluctuation remains   high.    However, both   the amplitude   and the velocity can grow with time leading to the instability when $\nu_0$ exceeds some critical value $\nu_c$ (see the dash-dotted lines in subplots (a) and (c)). This critical value may differ  due to a different set of  other parameter values. Here, the instability occurs with positive amplitude of shocks. However, there may be some parameter regime at which the collision frequency is relatively low but the effects of the  thermal feedback  to the density and temperature variations are typically high, e.g., $\Gamma=3,~\chi=0.8,~\nu_0=0.1,~q_{\rho_0}=0.4$ and $q_{T_0}=0.4$ at which  even though  $G(\tau)~(<0)$ decays, the velocity of shocks $d\phi/d\tau$ may increase with time. It follows that the rarefactive thermoacoustic shocks may exist   with an increasing speed and can achieve its maximum value at  finite distance and time before they damp away.       On the other hand, subplot (b) shows that the relative propagation distance at which the shocks may eventually damp away can be different for different values of $\Gamma$, $q_{\rho_0}$ and $q_{T_0}$. Here, one must note that  while the values of  both $\phi$ and $d\phi/d\tau$ increase,  those of $G(\tau)$ decrease with decreasing values of $\Gamma$ (not shown in the figure).
\par 
Figure  \ref{fig4-analytic}  shows the profiles of the shock solution \eqref{eq-sol-shock1} for different values of the parameters that characterize the thermal feedback  ($q_{\rho_0}$ and $q_{T_0}$), kinematic viscosity $(\mu)$, thermal diffusivity $(\chi)$, particle collision $(\nu_0)$, and the heat capacity $(\Gamma)$. We consider those parameter values for which       $|D|\lesssim|A|,~|B|$ (\textit{cf}.   Fig. \ref{fig2-ABD}) hold.   It is seen that depending on the parameter regime, not only do both the compressive and rarefactive shocks   appear, but    a transition from oscillatory to monotonic shocks and vice versa can also occur. Since they appear as double-layer shocks with multiple wave fronts \cite{misra2010},   both the free and trapped particles can adjust themselves at any time  to maintain the quasineutrality on each side of the propagating shocks.  From  subplot  (a)  of Fig.  \ref{fig4-analytic} it is evident that   as the value of $q_{\rho_0}$ is reduced, keeping the other parameters fixed, an oscillatory shock  with two wave fronts transits into  a monotonic one with a reduced wave amplitude (see the solid and dashed lines in subplot (a)). The shock profile remains monotonic in the interval $0.34\lesssim q_{T_0}\lesssim0.4$ and for fixed values of the other parameters, namely $q_{\rho_0}=0.16$, $\Gamma=3$, $\chi=0.8$, $\nu_0=0.1$ and $\mu=0$. In this regime, the monotonic shocks having amplitude $\sim0.04$ and velocity $\sim0.03$  will damp away after traveling a finite distance $\phi\sim0.4$ in time  $\tau=200$. Physically,    when the collision frequency is below its critical value, as the thermal feedback to the density fluctuation decreases, the magnitude of $D$ increases, i.e.,  the effect of the nonlinearity associated with the particle collision and the thermal feedback   is more pronounced and it may even dominate over the thermal diffusion and/or the kinematic viscosity. As a result, the oscillatory shocks transit into monotonic ones.
We also note that the shock wave amplitude increases with decreasing values of $q_{\rho_0}$. This is seen in the inverse relationship between $G(\tau)$ and $A$ in Eq. \eqref{eq-G-tau1} and the direct relationship between $A$ and $q_{\rho_0}$ for $\nu_0 \gtrsim 0.5$ observed in Fig. \ref{fig2-ABD}.
 Thus,  one can have a monotonic shock with reduced amplitude.
 Furthermore, such monotonic shocks again transit to oscillatory ones  with an increased amplitude when the value of $q_{T_0}$ is further reduced from $q_{T_0}=0.34$ to $q_{T_0}=0.165$ (see the dotted and dashed lines in subplot (a)). In this case,  the thermal diffusion and/or kinematic viscosity again dominate  over   the nonlinearity proportional to $D$ due to low thermal feedback of the medium to the temperature fluctuation. In this case, the wave amplitude increases due to the small effects of the collisional nonlinearity $(D)$ compared to the nonlinear convection $(A)$.   The profile  remains oscillatory in nature in the interval $0<q_{T_0}<0.34$ with $q_{\rho_0}=0.16$, $\Gamma=3$, $\chi=0.8$, $\nu_0=0.1$ and $\mu=0$. In this case, the oscillatory shocks having amplitude $\sim0.03$ and velocity $\sim0.02$ will damp in time $\tau=200$ after a finite distance  $\phi\sim0.5$.  
  However, the oscillatory shock can also transit into a monotonic one with an increasing value of the coefficient of the kinematic viscosity $\mu$ (See the solid and dash-dotted line). 
The effects of the parameters (namely, $\Gamma$, $\chi$ and $\nu_0$ that are  associated with  the heat capacity, the thermal diffusivity, and the particle collision respectively) on the profiles of shocks are also examined (subplot (b)). It is noted that, although the oscillatory nature of the shock profiles is preserved, the amplitude increases with a reduction of the heat capacity (see the solid and dashed lines). However, a transition from  an oscillatory to a monotonic structure  with reduced amplitude can also occur due to an enhancement of $\nu_0$ (below $\nu_c$) or      a decrement of    $\chi$ (see the dotted and dash-dotted lines).   
  On the other hand, subplot (c) of Fig.  \ref{fig4-analytic}  shows that the shock profile can also become rarefactive with negative amplitude for slightly a different set of values of the parameters that characterize either a reduction of the thermal feedback of the medium to the temperature variation or an enhancement of the same to the density variation.  These occur, respectively   when either $q_{\rho_0}$ is increased from $0.3$ to $0.4$ at fixed $q_{T_0}= 0.4$ or $q_{T_0}$ is decreased from $0.4$ to $0.28$ with fixed $q_{\rho_0}= 0.3$, all other parameters held constant. This is seen in the solid, dashed, and dotted lines of Figure \ref{fig4-analytic}(c).  It is also seen that as time goes on, the amplitude of the shock profile gets reduced (see the solid and dash-dotted lines) implying that  the shock wave may be damped due to a small effect of the collisional nonlinearity. However, as  mentioned before,  even though the amplitude decays, the velocity of shocks can increase with time  and achieve a maximum value before they damp  away.  Similar characteristics of shocks with variable velocity and decaying amplitude with time have been observed in incompressible fluids \cite{kedrinskii2001}.  
In the following section \ref{sec-numeric}, we will investigate the existence and characteristics of different kinds of shocks numerically and examine any qualitative agreement with the analytical results.  
 \subsection{Numerical approach}\label{sec-numeric}
The purpose of this section is to find a numerical solution of Eq. \eqref{eq-Burgers} and to verify whether this solution can match the analytic solution obtained in Sec. \ref{sec-analytic} for two different values of $\Gamma$: $\Gamma=2$ and $\Gamma=3$, keeping the other parameter values fixed at $q_{\rho_0}=0.27$, $q_{T_0}=0.34$, $\chi=0.8$, $\nu_0=0.1$, and $\mu=0$.  To solve Eq. \eqref{eq-Burgers} in time, we discretize the variable $\rho$ in both space  (where the first and second spatial derivatives are approximated by their respective second-order accurate central difference approximations (See Appendix \ref{sec-app-C}))  and time so that the solution is only defined  at the discrete points, separated by the spatial grid size  $\Delta\xi$ and the time step $\Delta\tau$.   We perform the simulation with the time interval $0\leq\tau\leq200$ and the space interval $-100\leq\xi\leq100$.   We use $N_{\tau}=2\times10^5$ intervals in time and $N_\xi=1000$ intervals in space such that  $\Delta\xi=0.01$ and $\Delta\tau=0.001$. In the numerical scheme, time is discretized as $\tau=\tau^i=i\Delta\tau$ with $i=0,1,2,...,N_{\tau}$ and space as $\xi=\xi_j=j\Delta\xi$ with $j=0,1,2,...,N_\xi-1$.      Also, we use periodic boundary conditions  for $\rho(\xi,\tau)$, i.e., $\rho(L_\xi,\tau)=\rho(0,\tau)$ (or for the discretized solution, $\rho^i_{N_\xi}=\rho_0^i$, where $\rho(\xi_j,\tau^i)\equiv\rho_j^i$ and $L_\xi$  is the length of the spatial interval) and its spatial derivatives. Next, we use the fourth-order Runge-Kutta scheme for the time-stepping with an initial   condition $\rho(\xi,0)=0.5[1+\tanh\left(\xi/8\right)]$. The time discretization of the Runge-Kutta scheme is given in Appendix \ref{sec-app-C}.   In the numerical integration, various spatial grid size and the time step combinations can be used. However, the most accurate results are obtained with $\Delta\xi=0.01$ and $\Delta\tau=0.001$ for which the $L_2$ error assumes a value $\lesssim10^{-3}$, where $L_2=\sqrt{\Delta\xi\sum^{N_\xi}_{i=0}|\rho^a-\rho^n|^2}$ with $\rho^a$ and $\rho^n$ denoting the solutions obtained by analytical and numerical methods.  It can be shown by von Neumann's stability analysis that the present scheme is also unconditionally stable \cite{banaja2015}. 
\par
\begin{figure*}
\centering
\includegraphics[width=6in,height=3.0in]{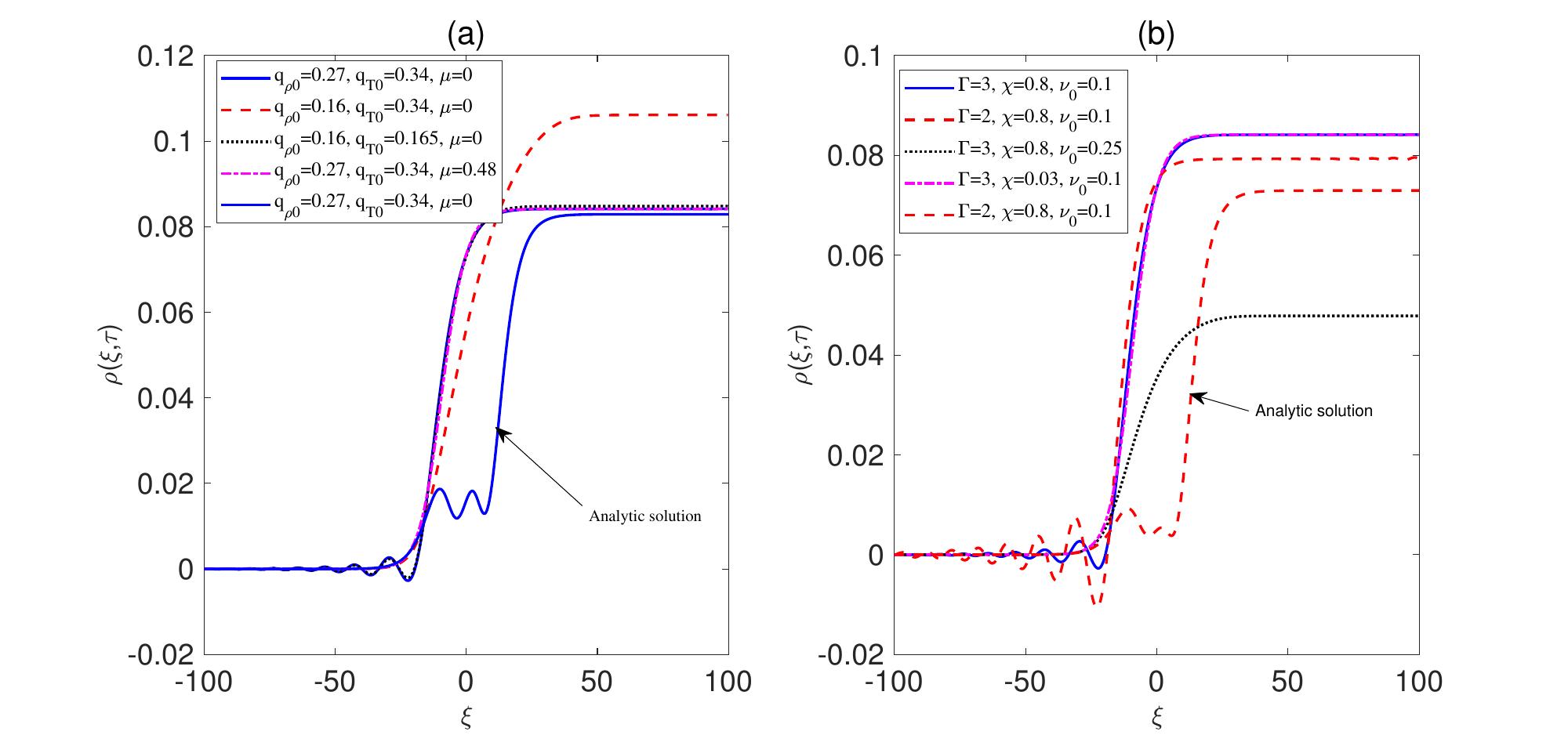}
\caption{The development of an initial   profile  $\rho(\xi,0)=0.5[1+\tanh\left(\xi/8\right)]$  into shocks (Numerical solution of Eq. \eqref{eq-Burgers})   is shown after $\tau=200$ for different sets of parameter values as in the legends and/or   in Fig. \ref{fig4-analytic}. As a comparison, the analytic solution \eqref{eq-sol-shock1} is plotted along with the numerical solutions of Eq. \eqref{eq-Burgers}  in two particular regimes  of $\Gamma$:  $\Gamma=3$ (subplot (a)) and $\Gamma=2$ (subplot (b)). Unless stated otherwise, the parameters values are $q_{\rho_0}=0.27$, $q_{T_0}=0.34$, $\chi=0.8$, $\nu_0=0.1$, and $\mu=0$.  }
\label{fig5-numeric}
\end{figure*}

 
  For the simulation, we consider the same parameter values as for Fig. \ref{fig4-analytic} and consider the pulse size $L_p<L_\xi$ so that the shock solutions exist.     The results are displayed in Fig. \ref{fig5-numeric}. A good qualitative agreement (i.e., overall trends like how $\rho$ generally depends on $\xi$) between the numerical results and those obtained from the analytical solutions is noted except near $\xi=0$ and in the asymptote ($\rho=0$) of the shock profile where some oscillatory nature of the analytic solution is found.  As an illustration, we have plotted the analytic solution along with the numerical solution of Eq. \eqref{eq-Burgers} in two particular cases as shown in subplots (a) and (b).  Initially, the wave steepens with growing amplitude, however, as time progresses, it evolves into a steady structure with wave fronts behind the shocks. The amplitude of the shocks so formed decays with time due to a small effect of the nonlinear  term proportional to $D$. It is noted that the wave fronts behind the shock are ordered and the one with the maximum amplitude is nearest to the shock.  From subplot (a) it is found that similar to Fig. \ref{fig4-analytic}, the wave steepening occurs  with a small reduction of the value  of $q_{\rho_0}$ and as a result, the oscillatory shock  transforms  into a monotonic one with an increased amplitude   (See the solid and dashed lines). The   monotonic profile again transits back into the oscillatory one with almost the same profile as for the increased values of $q_{\rho_0}$ and $q_{T_0}$ (solid line) with a small reduction of $q_{T_0}$ (See the dotted line). The similar qualitative features  (in agreement with subplot (b) of Fig. \ref{fig4-analytic})  also occur by the effects of the other parameters, namely $\Gamma$, $\chi$ and $\nu_0$. It is to be noted that for a different set of parameter values at which the coefficients $A$, $B$ and $D$ of Eq. \eqref{eq-Burgers} are significantly high or $D$ becomes larger than or comparable to  $A$ and $B$, the shock solution evolves with growing amplitude and may eventually blow up after a finite time. This happens when the collision frequency exceeds a critical value and the values of $q_{\rho_0}$ and/or $q_{T_0}$ are relatively high. 
\section{Conclusions}
The linear and nonlinear theories of plane progressive thermoacoustic waves are studied in multi-fluid complex plasmas with the effects of the thermal force due to a temperature gradient, the particle collision, and the kinematic viscosity.  In the linear regime, it is shown that in contrast to typical ion-acoustic waves in plasmas, the thermoacoustic waves in complex plasmas can be dispersionless both for long- $(k\ll1)$ and small-wavelength $(k\gtrsim1)$  oscillations. They can propagate at a near-acoustic speed in a wide range of parameter values (that are associated with the particle collision and the thermoacoustic interactions) as the wavelength approaches the plasma Debye length. Furthermore, in absence of the effects of the kinematic viscosity, a purely growing or damped thermoacoustic mode can exist when the collision frequency is above or below a critical value. However, the instability growth rate can be diminished when the effects of  either the particle collision   or the   kinematic viscosity are more pronounced.    
\par 
Although the onset of thermoacoustic instability in complex plasmas was reported in previous studies in the linear regime \cite{kryuchkov2018}, its consequences in the nonlinear regime have not been studied before in complex plasmas.  Recently, the linear theory of thermoacoustic wave amplification as well as the generation of thermoacostic shock waves have been reported, however, in different contexts, e.g.,  in a minimal-unit looped resonantor with the simulation of Navier-Stokes equations of compressible gases  \cite{gupta2017} and in a gas column with a high temperature gradient \cite{biwa2011,biwa2014}.  In the present investigation, starting from a set of fluid equations for viscous complex plasmas and using the reductive perturbation technique,  we have shown that the evolution of plane progressive thermoacoustic shocks can be described by the Burgers-like equation with  an additional nonlinear term that appears due to the particle collision  and nonreciprocal interactions of
charged particles providing the thermal feedback. It is found that a transition from oscillatory to monotonic shocks  can occur when the effect of the kinematic viscosity becomes significant compared to that of the thermal diffusion. Furthermore,   the shock profiles are  not only of compressive type but they can be rarefactive  when either the thermal feedback to the density fluctuation is slightly enhanced or that to the thermal fluctuation is somewhat reduced.  It is  shown  that the positive thermal feedback of the medium to the density and temperature fluctuations can lead to both monotonic and oscillatory  double-layer shocks that can be damped due to a small effect of the  collisional nonlinearity associated with the    particle collision  and  the thermal feedback. However, with the effects of the positive thermal feedback when  the collision frequency exceeds a critical value, the shock wave amplitude can grow leading to the instability of thermoacoustic shocks.   Our analytical  approach agrees well with numerical simulations  of the qualitative characteristics   of the thermoacoustic shocks in complex plasmas. Although there is no direct observation of the existence of thermoacoustic shocks in complex plasmas, because of the physical analogy between the collective dynamics in complex plasmas (with nonreciprocal effective interactions, providing positive thermal feedback)  and that in chemical reactive media \cite{kryuchkov2018} as well as experimental observations of thermoacoustic shocks in compressible fluids and gases \cite{gupta2017,biwa2011,biwa2014}, we believe that the present results should be useful for designing new experiments in dissipative strongly coupled systems, dusty plasmas or complex plasma crystals  as well as atmospheric fluids where the positive thermal feedback, necessary for the thermoacoustic  instability, can occur due to inhomogeneities of  equilibrium density and pressure. Furthermore, the thermoacoustic shocks so formed can transport particles and can thereby accelerate them in  the medium. 
\appendix
\section{Reduction of the linear dispersion relation  and  scaling of new space and time coordinates    } \label{appendix-B}
Here, we reduce the dispersion relation \eqref{eq-disp}  in the form $\omega\simeq k+\delta\omega$ and   define different scaling of space and time in the reductive perturbation technique used in Sec. \ref{sec-nonl}. 
\par 
  The linear dispersion relation \eqref{eq-disp} can be rewritten as 
\begin{equation} \label{eq:B1}
\omega^2-k^2=-i \omega (\nu+\mu k^2)+ \frac{i q_{\rho} k^2 }{\omega+i \left( \chi k^2 + \frac{2 \nu}{\Gamma} -q_{T} \right)}.
\end{equation}
We note that while the terms proportional to $\nu$ and $q_\rho$ appear due to the particle collision and thermoacoustic interactions, the term proportional to $\mu$ corresponds to the fluid kinematic viscosity. In absence of these effects, Eq. \eqref{eq:B1} gives a usual acoustic mode $\omega=k$. So, when these effects are retained and considered as small, one can assume the thermoacoustic mode to be of the form $\omega \simeq k+\delta \omega$, where $\delta  \omega$ represents a small correction   to the  wave frequency due to the thermoacoustic interactions, the particle collision and the effects of kinematic viscosity.  In order to obtain an approximate expression for $\delta\omega$, we   substitute    $\omega=k$ on the right-hand side of  \eqref{eq:B1}, which gives
\begin{equation}
\omega^2-k^2=-i k (\nu+\mu k^2)+ \frac{i q_{\rho} k^2 }{k+i \left( \chi k^2 + \frac{2 \nu}{\Gamma} -q_{T} \right)}.
\end{equation}
Next, for smallness of the terms proportional to $\nu$, $\mu$ and $q_\rho$, we obtain  the following expression.  
\begin{equation} \label{eq:B0}
\frac{\omega}{k}=1- \frac{i}{2k} (\nu+\mu k^2)+ \frac{1}{2} \frac{i q_{\rho} }{k+i \left( \chi k^2 + \frac{2 \nu}{\Gamma} -q_{T} \right)}.
\end{equation}
Thus, from Eq. \eqref{eq:B0}, we have
\begin{equation} \label{eq:B2}
\delta \omega (k)=- \frac{i}{2} (\nu+\mu k^2)+ \frac{1}{2} \frac{i q_{\rho} k }{k+i \left( \chi k^2 + \frac{2 \nu}{\Gamma} -q_{T} \right)}.
\end{equation}
By separating the real and imaginary parts of Eq. \eqref{eq:B0}, one can   obtain, after a straightforward algebra, the relations \eqref{eq-omegaR} and \eqref{eq-omegaI}.  
\par
 In order to derive an evolution equation (of Burgers-type) for thermoacoustic shocks using the reductive perturbation technique, we define a set of new space and time coordinates that can be obtained using the dispersion relation \eqref{eq:B0} as follows. Typically, the  second order space derivative of a given field in Burgers equation   appears    due to the effects of  either the   thermal diffusion or the kinematic viscosity, or both of them. So,   to define new space and time scales in the reductive perturbation scheme, it is sufficient to disregard the collisional and the thermal correction terms in Eq. \eqref{eq:B0} for which one obtains $\omega=k-(i/2)\mu k^2$. The phase of a plane wave then becomes $kx-\omega t=k(x-t)+ (i/2)\mu k^2 t=\epsilon(x-t)+ (i/2)\mu \epsilon^2 t$,  if $k\sim \epsilon$. Thus, one can define a set of new space and time  coordinates as in Eq. \eqref{eq-stretch}.   
 \section{Expansion of $q$} \label{appendix-A}
We expand $q(\rho,T)$ about the equilibrium values $\rho=1$ and $T=1$     using the Taylor series expansion as follows:
\begin{align} \label{eqs:A1}
&q(\rho,T) = q(1+\tilde{\rho},1+\tilde{T}) \nonumber \\ 
&= q_0+(\tilde{\rho} q_\rho+\tilde{T}q_T)+(\tilde{\rho}^2 q_{\rho\rho}+\tilde{\rho}\tilde{T}q_{\rho T}+\tilde{T}^2 q_{TT})+\cdots,
\end{align}
where $\tilde{\rho}$ and $\tilde{T}$  denote the small deviations of $\rho$ and $T$  respectively  from the equilibrium state.  Also, 
  $q_\rho$, $q_{\rho \rho}$ etc.  denote, respectively,  the first order, second order  etc. derivatives of $q$ with respect to $\rho$   and $q_0=0$ is the  equilibrium value of the heat source at   $\rho=1$, $T=1$.   We assume $\tilde{\rho}=\varepsilon \rho_1 + \varepsilon^2 \rho_2+\cdots$ and  $\tilde{T}=\varepsilon T_1 + \varepsilon^2  T_2+\cdots$, and the smallness of the parameters as   $q_\rho=\varepsilon q_{\rho_ 0}$, $q_T=\varepsilon q_{T_0}$, $q_{\rho \rho}=\varepsilon^2 q_{\rho_0 \rho_0}$, $q_{TT}=\varepsilon^2 q_{T_0T_0}$, $q_{\rho T}=\varepsilon^2 q_{\rho_0 T_ 0}$ etc. in which $q_{\rho_ 0}$,   $q_{T_ 0}$ $q_{\rho_ 0\rho_0}$, $q_{T_ 0T_0}$ etc. are of the order of unity.  Thus, Eq. \eqref{eqs:A1} reduces to
\begin{align}
q(\rho,T) & =\varepsilon^2(q_{\rho_0} \rho_1 +q_{T_0} T_1)+\varepsilon^3(q_{\rho_ 0} \rho_2 +q_{T_0} T_2)+\cdots \nonumber \\
& = \varepsilon^2 q_1 +\varepsilon^3 q_2+\cdots,
\end{align}
where $q_1\equiv q_{\rho_0} \rho_1+q_{T_0} T_1$, $q_2\equiv q_{\rho_0} \rho_2+q_{T_0} T_2$ etc.
\section{Runge-Kutta time discretization} \label{sec-app-C}
In our numerical simulation scheme, we rewrite the Burgers equation \eqref{eq-Burgers} as
\begin{equation} \label{eq-Burg-app}
\frac{\partial \rho}{\partial \tau}=\frac{\partial}{\partial\xi}\left(-\frac{1}{2}A\rho^2+B\frac{\partial\rho}{\partial\xi}\right) +D \rho^2\equiv F(\rho,\xi,\tau).
\end{equation}
Here, the function $F$ contains the discretized variables where we use the difference approximations for the spatial derivatives, defined by,
\begin{equation}
\frac{\partial\rho(\xi_j)}{\partial \xi}\approx \frac{\rho_{j+1}-\rho_{j-1}}{2\Delta\xi},~\frac{\partial^2\rho(\xi_j)}{\partial \xi^2}\approx \frac{\rho^i_{j+1}-2\rho_j^i+\rho^i_{j-1}}{(\Delta\xi)^2}
\end{equation} 
for $j=0,1,2,...,N_{\xi}-2$. At the boundaries (for $j=0$ and $N_\xi-1$) we define the spatial derivatives as
\begin{equation}
\frac{\partial\rho(\xi_0)}{\partial \xi}\approx \frac{\rho_{1}-\rho_{N_\xi-1}}{2\Delta\xi},~\frac{\partial^2\rho(\xi_0)}{\partial \xi^2}\approx \frac{\rho^i_{1}-2\rho_0^i+\rho^i_{N_\xi-1}}{(\Delta\xi)^2},
\end{equation} 
 \begin{equation}
 \begin{split}
&\frac{\partial\rho(\xi_{N_\xi-1})}{\partial \xi}\approx \frac{\rho_{0}-\rho_{N_\xi-2}}{2\Delta\xi},\\
&\frac{\partial^2\rho(\xi_{N_\xi-1})}{\partial \xi^2}\approx \frac{\rho^i_{0}-2\rho_{N_\xi-1}^i+\rho^i_{N_\xi-2}}{(\Delta\xi)^2},
\end{split}
\end{equation} 
where we denote the unknowns as $\rho=\left[\rho_0,\rho_1,\rho_2,...,\rho_{N_\xi-1} \right]$ with $\xi=\left[\xi_0,\xi_1,\xi_2,...,\xi_{N_\xi-1} \right]$. The Runge-Kutta scheme then reduces to
\begin{itemize}
\item $R_1\leftarrow F\left(\rho,\tau\right)$
\item $R_2\leftarrow F\left(\rho+\Delta\tau R_1/2,\tau+\Delta\tau/2\right)$
\item $R_3\leftarrow F\left(\rho+\Delta\tau R_2/2,\tau+\Delta\tau/2\right)$
\item $R_4\leftarrow F\left(\rho+\Delta\tau R_3,\tau+\Delta\tau\right)$
\item $\rho \leftarrow \rho+\left(\Delta\tau/6\right)\left[R_1+2\left(R_2+R_3\right)+R_4\right]$.
\end{itemize} 
The last item gives the solution $\rho$ at time $\tau+\Delta\tau$, and the above five steps are repeated with new values of $\rho$ until we get the desired result or reach the end of the simulation.

 \section*{Declaration of Competing Interest}
The authors declare that they have no known competing financial interests or personal relationships that could have appeared to influence the work reported in this paper.
\section*{Acknowledgments}
 The authors wish to thank the anonymous referees for their constructive comments and valuable advice, which improved the manuscript significantly.   This work was partially supported by the SERB (Government of India) sponsored research project with sanction order no. CRG/2018/004475 and initiated and significant part completed when G. Banerjee was a Dr. D. S. Kothari Post Doctoral Fellow of the University Grants Commission (UGC), Govt. of India, with Ref. No. F.4-2/2006(BSR)/MA/18-19/0096), in the Department of Mathematics, Siksha Bhavana, Visva-Bharati University, Santiniketan-731 235, India.
\bibliographystyle{apsrev4-1}
\bibliography{Reference-TAS}

\end{document}